# Capstone Experiences in Developing Augmented Reality Tables for Community Organizations

H. Keith Edwards[1], Michael R. Peterson[1], and Francis Cristobal`[1]
[1]Department of Computer Science, University of Hawaii at Hilo, Hilo, Hawaii, United States

**Abstract** – *This paper examines two senior capstone experiences developed as augmented reality tables over the past two years. Both projects were public facing efforts that required working implementations. The first project was deployed at an astronomy center and focused on interactions between land use and ecological aspects of Hawaii's Big Island while the second project focused more on historical sites on the same island. Both projects leveraged brownfield development and existing code bases to allow for student success in spite of the impacts of the COVID-19 pandemic.*

## 1 Introduction

Capstone experiences in computer science provide students with an opportunity to use the skills that they have gained over the course of their program of study. When designed and implemented correctly, these projects can provide students with a source of satisfaction as well as a showcase piece for their portfolio when they apply for professional jobs.

In selecting capstone projects, it is important to find opportunities that are meaningful to both students and product owners and that utilize current technology. Furthermore, it is important that students be able to achieve success with their projects.

In this paper, we discuss how Augmented Reality tables can provide students with a meaningful capstone experience as part of their undergraduate studies in a four-year computer science program. The paper illustrates this through two case studies that show successful implementation and community utilization of these augmented reality tables.

The rest of the paper is organized in the following manner. Section 2 discusses related work pertaining to Augmented Reality Tables, software engineering methods applicable to the development of this technology, and the undergraduate capstone experience at four-year universities. Section 3 discusses two successful implementations of this technology by undergraduate computer science students deployed at the Imiloa Astronomy Center in Hilo, Hawaii and at the Kō Education Center in Honokaa, Hawaii. Section 4 examines conclusions that can be drawn from these case studies, whilst Section 5 looks at future work related to this research.

## 2 Related Work

This section of the paper discusses the related work. We begin by looking at augmented reality and its implementations in table format. Next, we look at software engineering development life cycles with a particular focus on agile methods for student projects. Finally, we discuss features pertaining to undergraduate student capstone project courses and the difficulties inherent in the selection of quality projects for these courses.

### 2.1 Augmented Reality

Gangurde defines augmented reality as *"a technology that supplements the real world with virtual information that appears to coexist in the same space as the real world* [1]." While this serves as a basic definition, augmented reality in fact exists on a mixed reality continuum with many different types. In this subsection, we briefly describe the mixed reality continuum, introduce some common augmented reality types, and discuss implementation of augmented reality tables.

#### 2.1.1 The Mixed Reality Continuum

Milgram and Kishino's reality-virtuality (RV) continuum was first introduced in 1994 [2].

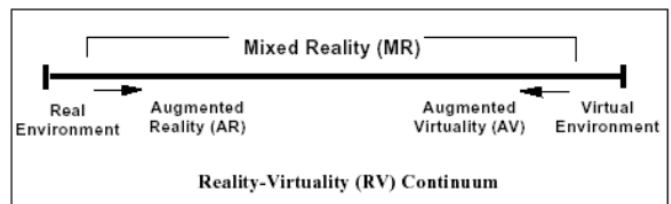

**Figure 1 – Reality – Virtuality (RV Continuum)**

The continuum shows real environments on the left end of the spectrum and virtual environments on the opposite end of the spectrum. Mixed reality exists between these two extremes. Mixed reality systems that are more based in reality are known as augmented reality (AR) whilst those more rooted in virtual environments are described as augmented virtuality (AV).

While various authors have extended or modified this existing framework in order to more accurately place work within the continuum [3], Gangurde's definition fits well within this model, and the work in this paper certainly falls within the AR segment of the continuum since users of Augmented Reality Tables do not immerse themselves in virtual environments in order to use the application.

### 2.1.2 Augmented Reality Applications and Forms

Augmented reality applications can take on a variety of different forms. One of the most well-known is the location based augmented reality game, Pokémon Go. The game was first launched in July 2016 and was the most popular and revenue generating mobile game at the time. Players loved the movement, sociability, and game mechanics that came as a part of the augmented reality experience [4]. Other successful examples of augmented reality include makeup applications from L'Oréal and IKEA's place application that allows you to see how different IKEA furniture pieces would look in your living space.

However, not all augmented reality applications are successful as evidenced by Google Glass, which was a small, lightweight wearable computational device akin to glasses. It featured a transparent display that allowed for hands-free work.

Early research focused on novel applications such as accessible web browsing for the visually impaired [5] and their use in wet-laboratory work [6]. However, the wearable nature of the device and its omnipresence in social situations led to privacy concerns and possible information overload [7]. Furthermore, the first round of adopters was a fairly elite group selected by Google, who had $1500 to pay for the device [8]. Hence, they became a symbol of a class divide within the San Francisco Bay Area and were not welcomed in a number of establishments. Adopters were frequently referred to as "glassholes" [8].

### 2.1.3 Augmented Reality Table Implementations

Augmented reality tables typically project virtual overlays from above on a physical surface. As users interact with that surface, the AR table updates the virtual overlay in real time in response to users' actions. Rogers et. al. describes an AR table system that overlays triggers upon a foosball table to provide semi-automatic game balancing for players of disparate skill levels. The AR-enhanced foosball table provided an enjoyable game experience even when matching novice and expert players [9].

Not every AR table system projects tables from above. The AR table system developed by Lam et. al. provides an augmented reality surface for playing physical trading card games such as "Magic: The Gathering" or "YU-GI-OH". The playing surface consists of a horizontal plasma TV for drawing information and generating sound, while an overhead camera system perceives card inputs and player commands. [10].

Regardless of the method of displaying information, AR Tables share several common traits: the ability to augment the physical world with virtual data, the capability of perceiving changes to the physical world through a computer vision system, and the ability to implement those changes in real time in response to user inputs.

## 2.2 Software Engineering Development Methods

The term software engineering was first coined in 1968 at a conference to discuss the software crisis brought on by the increased capabilities of computing machinery at the time to support the development of more complicated software. Since that initial conference, the field has continued to mature and to develop a number of techniques that assist in the development of quality software.

In particular, the field has developed a number of different software process models. A software process is a set of activities that leads to the production of a software product [11]. There are a number of different software processes such as the waterfall model [12], the Spiral Model [13], and the component-based model [14].

Since the early 2000's, rapid software development processes have dominated the conversation, with a particular focus on agile methods, which often feature direct customer engagement and accelerated delivery of incremental functionality to the customer at the expense of formal management procedures and extensive documentation [15]. Observationally, students enjoy working with agile methods due to increased engagement with the product owner(s) and a decreased focus on documentation.

Finally, reuse based software engineering can increase project success through techniques such as design patterns and application system reuse [16]. There is also brownfield software development, which generally refers to the development and deployment of new systems that need to interact with legacy systems. Brownfield development encourages reuse of existing software systems and can provide a starting point for development [17]. The AR Table projects fall under the reuse and brownfield umbrella since they have an existing code base, but new development needs to interface with that existing system.

## 2.3 Undergraduate Capstone Experiences

Capstone Courses are designed to culminate the undergraduate experience by allowing student to focus on a semester or yearlong project [18]. Capstone experiences also allow students to reflect upon what they have learned throughout the course of their undergraduate experience and to focus on applying those skills and methods most practicable to the project. The computer science curriculum at the University of Hawaii at Hilo requires a two-semester set of software engineering courses that form this capstone experience as part of the applied learning experience at the university. The course sequence utilizes a hybrid of a studio-based approach and the directed sponsored approach where students work with a product owner to develop a working piece of software and utilize the course meeting times for development [19].

As described in Adams, et. al., there are numerous problems associated with capstone experiences in software engineering [20]. Three notable challenges in UH-Hilo's Software Engineering sequence are the selecting of client proposed projects that align with student interests and skills, the selection of projects that utilize current technology, and the defining of projects so that students achieve success in the course.

In this regard, Augmented Reality Table implementations meet each of these three challenges. First, the tables utilize local data for Hawaii island (a.k.a. 'The Big Island'), so the project aligns with student interests as well as providing placed based science education [21]. Furthermore, the Augmented Reality Tables utilize skills that students have either previously developed or that are useful in industry, notably [Angular JS, Python, JSON, Google Earth, GIS data, etc.]. Finally, students can use the previously implementations as reference points, so there is a greater likelihood of project success.

# 3 Case Studies

In this portion of the paper, we examine the implementation of two augmented reality tables that were successfully constructed and implemented as part of the software engineering capstone experience in the University of Hawaii at Hilo's computer science department.

The first AR Table project was implemented to help visualize critical infrastructure and important land use on the Big Island of Hawaii. The project was implemented at the Imiloa Astronomy Center in Hilo, HI during the 2019-2020 Academic year.

The second AR Table project was a similar undertaking (i.e. brownfield development) that was implemented to help visualize important cultural landmarks on the Hamakua Coast along the northeast shore of the Big Island. It is currently located at the Kō Education Center in Honokaa, Hawaii.

## 3.1 Overall Development Methodology

The capstone course sequence in software engineering at the University of Hawaii at Hilo is a yearlong course that allows senior students to develop a working piece of software for a real-world product owner.

The two-course sequence uses as an agile software development methodology with approximately 10-12 separate sprints throughout the year, which culminate in a final product delivered at the end of the spring semester. Since the courses are focused on agile development, there is a requirement for product owners maintain close communication with their student development groups. Ideally, product owners should meet with their development teams at the end of each sprint in order to evaluate their development efforts and to provide feedback as to the direction of the product.

During the 2020-2021 academic year, the courses focused on software reuse coupled with brownfield development. This focus on reuse allowed for the reuse of existing code bases from the University of Hawaii at Manoa and the 2019-2020 project that had been previously developed. This code reuse helped to reduce the risk associated with the individual projects.

## 3.2 Imiloa Project Description

Imiloa Astronomy Center is a world class 40,000 square foot museum and planetarium. It is a unit of the University of Hawaii at Hilo located above the University of Hawaii Technology park in Hilo, Hawaii. It showcases the astronomical research at the summit of Mauna Kea and its connections to the rich traditions of Hawaiian culture. Imiloa means "to seek far" and offers a variety of educational programming, scientific exhibits for Pre-K-12 students and visitors year-round. Imiloa has a mix of showcases from discoveries from the space observatories on Mauna Kea and mastery of Polynesian sea navigation and wayfinding.

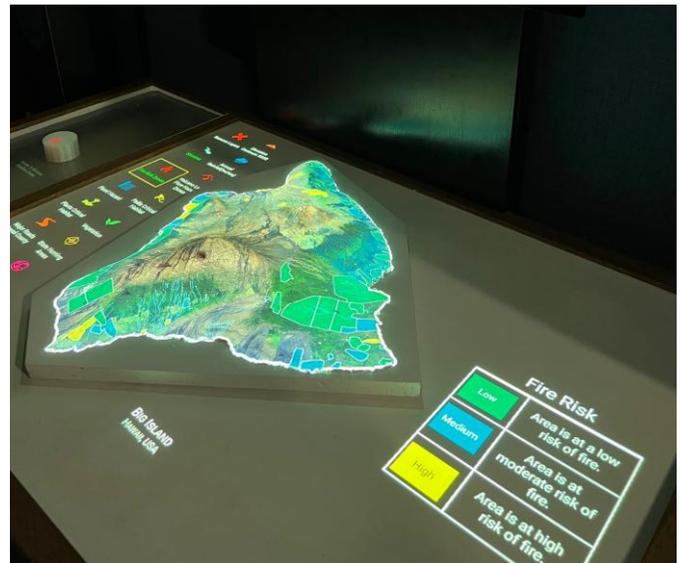

**Figure 2 – Big Island Fire Hazard Highlighted on Imiloa AR table**

The augmented reality table developed for Imiloa focuses on projecting spatial GIS data onto a 3D model of the Big Island. This implementation of the project had 15 different layers. Certain layers focused on geographical features of the island including elevation contours, streams, annual rainfall, fire risk zones, volcanic hazard zones, and flood hazard zones. Another set of layers focused on ecological aspects of the island including critical habitats for birds, critical plant areas, and vegetation. The next set of layers focus on land use including major roads, hunting areas, trails and tax parcels. A final layer shows the Ahupua'a, which are the traditional Hawaiian districts for the land, which ran from ocean to mountain top.

These multiple layers allow users to see the complex ecological and land use interactions that take place on the Big Island of Hawaii. For example, users can add multiple layers in order to examine the overlap between fire risk areas and native bird habitats.

The two-student team modified existing TypeScript, HTML and JavaScript files that were available from the original Oahu model implemented at UH Manoa. The students worked on the software while Imiloa outsourced the printing of the 3D Map of the Big Island as well as the construction of the exhibit table. The project was successfully implemented in spite of the pandemic disrupting instruction from March 2020 to May 20202. The completed AR Table is currently housed at Imiloa.

### 3.3 Kō Education Center Project Description

The Kō Education Center is a unit of Hawaii Community College focused on providing access to education in the North Hawaii community. It is located in Honokaa, Hawaii. It has a conference center as well as a Heritage center showcasing archives and exhibits from the local community. Kō Education Center offers college courses for the North Hawaii community. It is affiliated with the University of Hawaii system.

The AR Table for the Kō Education Center took place during the 2020-2021 academic year and involved three senior computer science students. Since COVID-19 had already impacted the previous semester, instruction and deliverables were able fully virtual during this project.

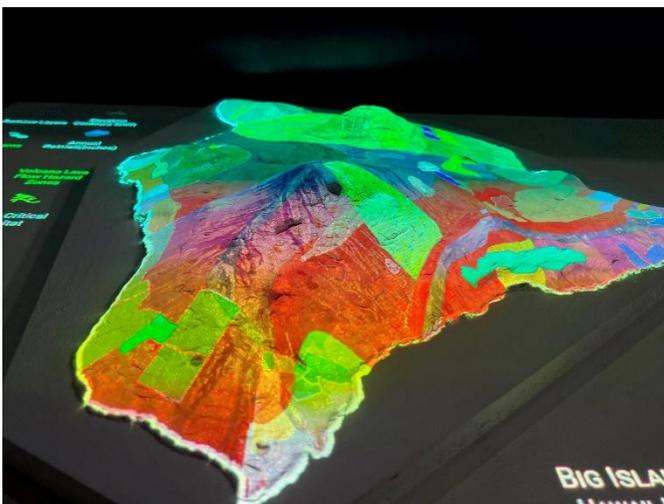

**Figure 3 – Layers Displayed on a 3D Printed Map of the Big Island**

The change in modality was offset by the fact that students were working with an existing code base that had been modified the previous year and were somewhat used to virtual instruction at this point.

The Kō Education Center project focused on adding additional layers of functionality to show cultural artifacts on the Hamakua Coast. Like the Imiloa implementation, the technology stack consisted of Python, Angular JS, JSON, and HTML as depicted in Figure 4 below.

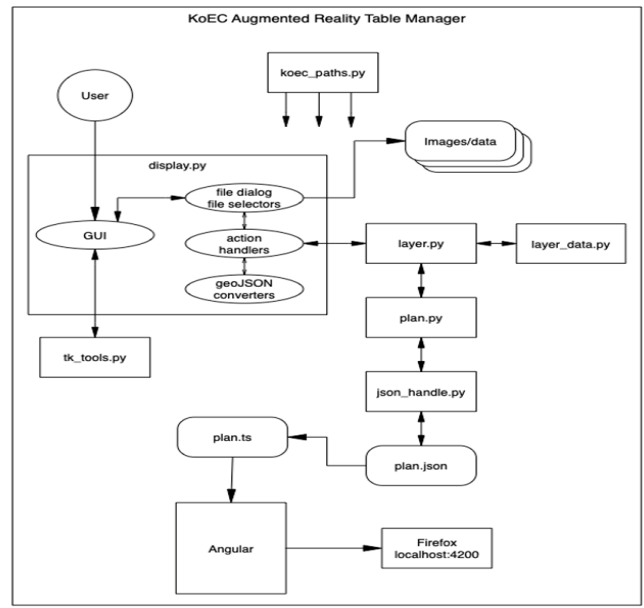

**Figure 4 – Technology Stack and Interfaces for Kō Education Center Project.**

The construction of the physical aspects for this table differed with the Kō Education Center taking the lead in 3D printing the topographic model of the Big Island, and students from Honokaa High School constructing the Physical table. Like its counterpart, it was successfully implemented, and students had a positive experience with the overall development.

## 4 Conclusions

Senior Capstone experiences are difficult undertakings for undergraduate students due to their extended nature and the requirement that they be fully implemented. In addition, many of these projects are public facing which adds pressure to implement working code.

This paper explored the development of two augmented reality tables undertaken by two different student project teams. The project at Imiloa focused on interactions between land use and ecological aspects of the Big Island while the Kō Education Center project focused more on historical aspects present on the Hamakua Coast.

Allowing students to use existing code bases and undertake brownfield development allows for a mitigation of the risks inherent in these undertakings. In particular, augmented reality tables possess these properties, and two separate groups of senior undergraduate students were able to make successful and highly visible public projects using this technology. This was in spite of the fact that both projects were impacted by the COVID-19 pandemic.

## 5   Future Work

The two projects featured in the case study resulted in highly successful software engineering capstone experiences for students. For future work, we would like to collect qualitative and quantitative data on the impact of the project implementations and the use of these devices in their public settings.

## 6   References

[1]   Madhavi Gangurde. 2011. Augmented reality. In Proceedings of the International Conference & Workshop on Emerging Trends in Technology (ICWET '11). Association for Computing Machinery, New York, NY, USA, 1363. DOI: https://doi.org/10.1145/1980022.1980349

[2]   MILGRAM, P., AND KISHINO, F. 1994. Augmented reality: A class of displays on the reality-virtuality continuum. In SPIE, Telemanipulator and Telepresence Technologies, vol. 2351, 42–48.

[3]   Skarbez Richard, Smith Missie, Whitton Mary C. Revisiting Milgram and Kishino's Reality-Virtuality Continuum. Frontiers in Virtual Reality, Vol 2. 2021. ISSN 2673-4192. DOI=10.3389/frvir.2021.647997

[4]   Janne Paavilainen, Hannu Korhonen, Kati Alha, Jaakko Stenros, Elina Koskinen, and Frans Mayra. 2017. The Pokémon GO Experience: A Location-Based Augmented Reality Mobile Game Goes Mainstream. In Proceedings of the 2017 CHI Conference on Human Factors in Computing Systems (CHI '17). Association for Computing Machinery, New York, NY, USA, 2493–2498. DOI: https://doi.org/10.1145/3025453.3025871

[5]   Najd A. Al-Mouh and Hend S. Al-Khalifa. 2015. Towards accessible web browsing for visually impaired people using Google Glass. In Proceedings of the 17th International Conference on Information Integration and Web-based Applications & Services (iiWAS '15). Association for Computing Machinery, New York, NY, USA, Article 12, 1–3. DOI: https://doi.org/10.1145/2837185.2837265

[6]   Grace Hu, Lily Chen, Johanna Okerlund, and Orit Shaer. 2015. Exploring the Use of Google Glass in Wet Laboratories. In Proceedings of the 33rd Annual ACM Conference Extended Abstracts on Human Factors in Computing Systems (CHI EA '15). Association for Computing Machinery, New York, NY, USA, 2103–2108. DOI: https://doi.org/10.1145/2702613.2732794

[7]   Qianli Xu, Michal Mukawa, Liyuan Li, Joo Hwee Lim, Cheston Tan, Shue Ching Chia, Tian Gan, and Bappaditya Mandal. 2015. Exploring users' attitudes towards social interaction assistance on Google Glass. In Proceedings of the 6th Augmented Human International Conference (AH '15). Association for Computing Machinery, New York, NY, USA, 9–12. DOI: https://doi.org/10.1145/2735711.2735831

[8]   Honan, Mat (May 15, 2013). "I, Glasshole: My Year With Google Glass". WIRED. Retrieved March 6, 2019.

[9]   Katja Rogers, Mark Colley, David Lehr, Julian Frommel, Marcel Walch, Lennart E. Nacke, and Michael Weber. 2018. KickAR: Exploring Game Balancing Through Boosts and Handicaps in Augmented Reality Table Football. In Proceedings of the 2018 CHI Conference on Human Factors in Computing Systems (CHI '18). Association for Computing Machinery, New York, NY, USA, Paper 166, 1–12. DOI: https://doi.org/10.1145/3173574.3173740

[10]   Albert H. T. Lam, Kevin C. H. Chow, Edward H. H. Yau, and Michael R. Lyu. 2006. ART: augmented reality table for interactive trading card game. In Proceedings of the 2006 ACM international conference on Virtual reality continuum and its applications (VRCIA '06). Association for Computing Machinery, New York, NY, USA, 357–360. DOI: https://doi.org/10.1145/1128923.1128987

[11]   Sommerville, I. (2016) Software Engineering. 10th Edition, Pearson Education Limited, Boston.

[12]   Royce, Winston (1970), "Managing the Development of Large Software Systems", Proceedings of IEEE WESCON, 26 (August): 1–9

[13]   B Boehm. 1986. A spiral model of software development and enhancement. SIGSOFT Softw. Eng. Notes 11, 4 (August 1986), 14–24. DOI: https://doi.org/10.1145/12944.12948

[14]   Ivica Crnkovic, Brahim Hnich, Torsten Jonsson, and Zeynep Kiziltan. 2002. Specification, implementation, and deployment of components. Commun. ACM 45, 10 (October 2002), 35–40. DOI:https://doi.org/10.1145/570907.570928

[15]   Beck, K., et al. (2001) The Agile Manifesto. Agile Alliance. http://agilemanifesto.org/

[16]   Hefedh Mili, Ali Mili, Sherif Yacoub, and Edward Addy. 2001. Reuse-based software engineering: techniques, organization, and controls. Wiley-Interscience, USA.

[17]   David G. Hill. 2008. Review of "Eating the IT Elephant: Moving from Greenfield Development to Brownfield (1st ed.)," IBM Press, 2008, $29.99, ISBN: 0137130120.: Review of "Eating the IT Elephant: Moving from Greenfield Development to Brownfield (1st ed.)," IBM Press, 2008, ISBN: 0137130120. Queue 6, 5 (September 2008), 59. DOI: https://doi.org/10.1145/1454456.1454469

[18] Richard Johnson, Peter C. Isaacson, Noel F. Lejeune, and Tim Reeves. 2002. Senior project/capstone courses. J. Comput. Sci. Coll. 18, 1 (October 2002), 184.

[19] N. Hari Narayanan, Christopher Hundhausen, Dean Hendrix, and Martha Crosby. 2012. Transforming the CS classroom with studio-based learning. In Proceedings of the 43rd ACM technical symposium on Computer Science Education (SIGCSE '12). Association for Computing Machinery, New York, NY, USA, 165–166. DOI: https://doi.org/10.1145/2157136.2157188

[20] Liz Adams, Mats Daniels, Annegret Goold, Orit Hazzan, Kathy Lynch, and Ian Newman. 2003. Challenges in teaching capstone courses. SIGCSE Bull. 35, 3 (September 2003), 219–220. DOI: https://doi.org/10.1145/961290.961575

[21] Semken, S. and Freeman, C.B. (2008), Sense of place in the practice and assessment of place-based science teaching. Sci. Ed., 92: 1042-1057. https://doi.org/10.1002/sce.20279